\documentclass[aps, prl, superscriptaddress,twocolumn, longbibliography]{revtex4-2}
\usepackage{soul}
\usepackage{graphicx}% Include figure files
\usepackage{bm}% bold math
\usepackage[normalem]{ulem}
\usepackage{amsbsy,amsmath}
\usepackage{soul}

\usepackage{palatino}
\usepackage[utf8]{inputenc}
\usepackage{epsf}
\usepackage{amsmath,amssymb}
\usepackage{calc}
\usepackage{amsmath,amssymb, bm}
\usepackage{float}

\usepackage[usenames,dvipsnames]{xcolor}

\usepackage{graphicx}
\usepackage{hyperref}

\newcommand{\ben}{\begin{equation}}
\newcommand{\een}{\end{equation}}
\newcommand{\bea}{\begin{eqnarray}}
\newcommand{\eea}{\end{eqnarray}}
\def\bra#1{\langle#1\vert}
\def\ket#1{\vert#1\rangle}

\def\sss{\scriptscriptstyle\rm}

\def\1s{_{1,\sss S}}
\def\2s{_{2,\sss S}}

\def\s{_{\sss S}}
\def\xc{_{\sss XC}}

\def\H{_{\sss H}}

\def\ext{_{\rm ext}}

\def\br{{\bf r}}

\begin{document}

\title{Harnessing dressed time-dependent density functional theory for the non-perturbative regime: Electron dynamics with double excitations
%Accessing double excitations in electron dynamics: Bootstrapping dressed time-dependent density functional theory to the non-perturbative regime 
}
%Strong field dynamics with double excitations: ... }
\author{Dhyey Ray}
\affiliation{Department of Physics, Rutgers University, Newark 07102, New Jersey USA}
%\affiliation{Department of Biomedical Engineering, Rutgers University, Piscataway, New Jersey USA}
\author{Anna Baranova}
\affiliation{Department of Physics, Rutgers University, Newark 07102, New Jersey USA}
\author{Davood B. Dar}
\affiliation{Department of Physics, Rutgers University, Newark 07102, New Jersey USA}
\affiliation{Department of Physical and Environmental Sciences University of Toronto, Canada}
\author{Neepa T. Maitra}
\affiliation{Department of Physics, Rutgers University, Newark 07102, New Jersey USA}
\email{neepa.maitra@rutgers.edu}

\date{\today}

\begin{abstract}
Recent progress has been made in capturing spectral features of electronic states of double-excitation character in time-dependent density functional theory (TDDFT) through a frequency-dependent kernel. 
While it might appear that this development is limited to the perturbative regime, we show that when used within response-reformulated TDDFT, it accurately captures strong-field dynamics involving states of double-excitation character. More generally, this demonstrates how RR-TDDFT enables exchange-correlation functional developments in the response regime, which have so far been more successful than those in the non-linear regime, to be exploited for non-perturbative dynamics, thus significantly broadening their range of applications.
\end{abstract}

\maketitle

%\section{Introduction}
While time dependent density functional theory (TDDFT)~\cite{RG84,CarstenBook,M16} has emerged as a method of choice for calculating electronic spectra and non-perturbative dynamics,  it is well-recognized that the approximate functionals used today suffer from well-known failures. These are particularly severe in the non-perturbative regime, as demonstrated by e.g., resonantly-driven dynamics~\cite{RB09,FHTR11,EFRM12,LFSEM14}, long-range charge-transfer~\cite{FERM13,M17,RN11,LM21}, and unphysical spurious peak-shifting~\cite{HTPI14,RN12c,FLSM15}.
One key reason is that the exchange-correlation (xc) potential, $v\xc[n;\Psi_0,\Phi_0](\br,t)$, that appears in the time-dependent Kohn-Sham (TDKS) equations of TDDFT depends on the history of the density, the initial many-body state, and the initial Kohn-Sham (KS) state, but this memory-dependence is neglected in the calculations: the commonly used adiabatic approximations depend only on the instantaneous density, $v\xc^{\rm A}[n;\Psi_0,\Phi_0](\br,t) = v\xc^{\rm g.s.}[n(t)](\br)$ where $v\xc^{\rm g.s.}[n](\br)$ is a ground-state functional. 
Practical memory-dependent functionals in the non-perturbative regime remain elusive, despite intense and varied efforts~\cite{LM23}.
Orbital-dependent functionals contain some degree of memory since instantaneous functionals of the orbitals contain some history-dependence of the density, but these have been limited to exact-exchange in hybrid functionals, and miss important memory effects in the correlation potential, such as step and peak features~\cite{LM23,DLM22}.

So far, there has been more success for memory-dependent functional development in the linear response regime, notably for double excitations~\cite{TAHR99,TH00,C05,MW09,RSBS09,MZCB04,DM23,DM25} and dissipation/relaxation in metallic systems~\cite{VK96,VUC97}. In this regime, memory-dependence manifests as frequency-dependence in the xc kernel, which is the Fourier transform of $f\xc[n^{\rm g.s.}](\br,\br',t-t') = \left.\frac{\delta v\xc[n](\br,t)}{\delta n(\br',t')}\right\vert_{n = n^{\rm g.s.}}$, generating shifts and creating new excitation energies with corrected transition densities. 

Recently,  a reformulation of TDDFT was proposed which offers the possibility to use these successes for fully non-perturbative dynamics~\cite{DBM24}. This Response-Reformulated TDDFT (RR-TDDFT) by-passes the xc potential of TDKS, solving instead for coefficients of the many-body wavefunction, without needing the wavefunction itself, and requires xc functionals only in the ground state, linear, and quadratic response regimes. Since the xc functionals are thus evaluated only near a ground state, adiabatic approximations perform far better than they do for the xc potential appearing in the TDKS equation, redeeming, for example, Rabi oscillations for which the traditional TDKS with adiabatic functionals fails~\cite{DBM24}. 
However, RR-TDDFT is limited by the accuracy of the xc functional in the response regime, and a well-known failure of the adiabatic approximation is the inability to capture excitations that have some double-excitation character (double excitations, {\it tout court})~\cite{TH00}. 
On the other hand, non-perturbative TDKS with adiabatic functionals can access double excitations, but the description is unreliable, unphysically dependent on the field parameters driving the system~\cite{IL08}.
Recently, a non-adiabatic ``dressed" xc kernel was derived that captures both  excitation energies and oscillator strengths of double-excitations~\cite{DM23,DM25}. While shown to successfully predict the 2$^1$Ag -- 1$^1$Bu curve-crossing and doubles-character in butadiene, a challenge even for high-level wavefunction methods~\cite{PSLP21}, and excited-state densities in model one-dimensional (1D) systems~\cite{BM25}, the prospect of using the kernel for fully non-linear processes beyond the response regime had not been exploited, let alone envisioned.

%%%%%%%% This was written before %%%%%%%%%%%
% In this work, {\color{magenta}for the first time} we show {\color{magenta} that the dressed TDDFT approach can be extended beyond the linear response regime, enabling access to the dynamics of the system driven to the doubly excited states.}
%%%%%%%%%%%%%%%%%%%%%%%%%%%%%%%%%%%%%%%%%%%%

%\sout{In this work, we extend dressed TDDFT beyond linear response applications via RR-TDDFT, making real-time dynamics involving double excitations achievable via TDDFT with as much reliability as the dressed approach for describing their excitations and response properties.}
In this work, we extend dressed TDDFT beyond its linear-response application by employing RR-TDDFT. This extension enables real-time simulations of non-perturbative dynamics involving double excitations with a reliability comparable to that of dressed TDDFT in describing their excitation energies and response properties. More generally, this work shows how the RR-TDDFT formalism enables any functional developed in the response regime to now be useful for fully non-equilibrium dynamics.

RR-TDDFT operates by solving a system of ordinary differential equations (ODEs) for coefficients of the many-body time-dependent state expanded in unperturbed many-body states $\Psi_m$,  in the flavor of time-dependent configuration interaction (TDCI), but with the important difference that no many-body states are actually needed or calculated. 
 We note that RR-TDDFT could be viewed as an exactification of the approach to strong-field molecular dynamics in Refs.~\cite{CL21,SS11b,MLR14b} where linear-response TDDFT quantities and auxiliary wavefunctions were used in a TDCI-like framework.
Under the Runge-Gross theorem~\cite{RG84}, with
%the same condition of usual TDDFT, stemming from the Runge-Gross theorem~\cite{RG84} that 
the applied potential being a one-body multiplicative operator, all the input ingredients to the ODEs, as well as all observables, can be obtained from ground-state DFT or response TDDFT: energies $E_m$ (from ground-state ($m=0$) and linear-response~\cite{PGG96,C95,C96,GPG00}), transition densities out of the ground state, $\rho_{0m}(\br) = \bra{\Psi_0}\hat{n}(\br)\ket{\Psi_m}$ (from linear response~\cite{PGG96,C95,C96,GPG00}), transition densities between excited states, $\rho_{mn}(\br) =\bra{\Psi_m}\hat{n}(\br)\ket{\Psi_n}$ (from quadratic response~\cite{PF18,SVHA02}), and state densities $\rho_{mm}(\br) = \bra{\Psi_m}\hat{n}(\br)\ket{\Psi_m}$ (from ground-state DFT ($m=0$) or linear response~\cite{F01,FA02,BM25}). The equations are:
\begin{equation}
i\,\dot{C}_m(t) = E_m\,C_m(t) + \sum_{n} V^{\mathrm{app}}_{mn}(t)\,C_n(t),
\label{eq:coeff-dyn}
%\tag{2}
\end{equation}
with the driving matrix elements defined as
\begin{equation}
V^{\mathrm{app}}_{mn}(t)
= \langle \Psi_m \lvert V^{\mathrm{app}}(t) \rvert \Psi_n \rangle
= \int d^{3}r\; v^{\mathrm{app}}(\mathbf{r},t)\,\rho_{mn}(\mathbf{r})
\label{eq:vapp}
%\tag{3}
\end{equation}
and observables extracted (in principle~\cite{RG84}) from the density
\ben
n(\br,t) = \sum_{n,m}C_n^*(t)C_m(t)\rho_{nm}(\br).
\label{eq:density}
\een
We will consider here uniform applied electric fields, $\bm{\mathcal{E}}(t)$, such that $V_{mn}^{\rm app}(t) = \bm{\mathcal{E}}(t)\cdot {\bf d}_{mn}$, and the dipole moment as our observable, $\mathbf{d}(t) = \sum_{n}\sum_{m} C_n^{*}(t)\,C_m(t)\,\mathbf{d}_{nm}$, where $\mathbf{d}_{nm} = \left\langle \Psi_n \left| \hat{\mathbf{r}}\right|\Psi_m \right\rangle = \int d^3 r \,\mathbf r \rho_{nm}(\mathbf r)$.
{ The xc functional (ground-state and response, as detailed above) enter in the calculation of the energies $E_m$, the transition densities $\rho_{mn}$ and the densities $\rho_{mm}$, both of which weigh the integral in the matrix element of the applied potential $V^{\rm app}_{mn}(t)$.}
%\begin{equation}
%i\,\dot{C}_m(t) = E_m\,C_m(t) + %\bm{\mathcal{E}}(t)\!\cdot\!\sum_{n}\mathbf{d}_{mn}\,C_n(t),
%\label{eq:coeff-dyn-dip}
%\tag{4}
%\end{equation}
%and the dipole expectation value is
%\begin{equation}
%\mathbf{d}(t) = \sum_{n}\sum_{m} %C_n^{*}%(t)\,C_m(t)\,\mathbf{d}_{nm}.
%\label{eq:dipole}
%\tag{5}
%\end{equation}

 The advantage of RR-TDDFT over traditional TDKS calculations becomes salient when considering the nature of the xc functionals that are needed in each approach: The TDKS equations, $\mathrm i \partial_t\phi_i(\br)=\left\{ -\frac12\nabla^2 + v\s(\br) \right\}\phi_i(\br)$ where $v\s(\br,t) = v\ext(\br,t) + v\H[n](\br,t) + v\xc[n,\Psi_0,\Phi_0](\br,t)$, require the xc potential evaluated in the fully non-equilibrium domain, but adiabatic approximations simply take a ground-state functional, $v\xc^{\rm A}[n,\Psi_0,\Phi_0](\br,t) = v\xc^{\rm g.s.}[n(t)](\br,t)$, approximating the xc effects of the fully non-equilibrium system by ground-state xc. On the other hand, as is evident in Eqs.~(\ref{eq:coeff-dyn})--~(\ref{eq:density}) above, the xc functional is required only near the ground state for RR-TDDFT, that is, near the domain for which the adiabatic functionals were derived. 
Ref.~\cite{DBM24} showed how Eqs.~(\ref{eq:coeff-dyn})--(\ref{eq:density}) were able to capture Rabi oscillations and charge-transfer dynamics using adiabatic approximations, even when those same adiabatic approximations  dramatically failed when used within the traditional TDKS scheme. 
% \sout{However, RR-TDDFT is only as good as the approximate functional it uses is in the response regime, and a well-known failure of the adiabatic approximation is the inability to capture states of double-excitation character (double-excitations, {\it tout court})~\cite{TH00}. Recently, a non-adiabatic ``dressed" xc kernel was derived, building on the earlier work of Ref.~\cite{MZCB04}, that captures both the excitation energy and oscillator strengths of double-excitations~\cite{DM23,DM25}. While it was shown to successfully predict the 2Ag-1Bu curve-crossing and doubles-character in butadiene, a challenge even for high-level wavefunction methods~\cite{PSLP21}, and excited-state densities in model one-dimensional (1D) systems~\cite{BM25}, the prospect of the kernel being used for fully non-linear processes beyond the response regime had not been exploited, let alone envisioned. }
% {\it may end up adjusting and moving to intro later}. 
However, 
%R-TDDFT is only as good as the approximate functional is in the response regime, and 
when the dynamics accesses states whose response properties are not well-predicted by adiabatic approximations, then RR-TDDFT with these functionals will also not perform well. This is the case for double excitations. Instead, the frequency-dependent kernel of dressed TDDFT (DTDDFT)~\cite{DM23} predicts their excitation energies, transition densities out of the ground-state, and densities~\cite{BM25} well. 
We now use the DTDDFT xc kernel in RR-TDDFT to predict fully non-perturbative dynamics that access double-excitations. 

% Before doing this, we briefly recall  the TDDFT response framework from which we will extract the ingredients needed for RR-TDDFT evolution.  
Excitation energies in response TDDFT are obtained as eigenvalues of the matrix
%\begin{equation}
%\Omega(\omega)\, \mathbf{G}_I = %\omega_I^{2}\, \mathbf{G}_I \,,
%\tag{6}
%\end{equation}
%where
\begin{equation}
\Omega_{qq'}(\omega) = \nu_q^{2}\,\delta_{qq'} 
+ 4\,\sqrt{\nu_q \nu_{q'}}\, f_{\mathrm{HXC},{qq'}}(\omega)
%\tag{7}
\label{eq:casida}
\end{equation}
with $f_{\mathrm{HXC},{qq'}}(\omega) = 
\iint d\mathbf{r}\, d\mathbf{r}' \,
\Phi_q(\mathbf{r}) \, f_{\mathrm{HXC}}(\mathbf{r},\mathbf{r}',\omega)\,
\Phi_{q'}(\mathbf{r}')$ and 
$f_{\mathrm{HXC}}(\mathbf{r},\mathbf{r}',\omega) = 
\frac{1}{|\mathbf{r}-\mathbf{r}'|} + f_{\mathrm{XC}}(\mathbf{r},\mathbf{r}',\omega)$.
Here the KS transition density $\Phi_q(\mathbf{r}) = \phi_i^*(\mathbf{r}) \phi_a(\mathbf{r})$ is the product 
of occupied ($i)$ and unoccupied ($a$) KS orbitals, and $\nu_q = \epsilon_a - \epsilon_i$ are their orbital energy differences. The oscillator strengths and transition densities are 
 obtained from eigenvectors $\mathbf{G}_I$ of  the matrix Eq.~(\ref{eq:casida})
 when normalized according to~\cite{C95}
\begin{equation}
\mathbf{G}_I^\dagger 
\left( 1 - \left.\frac{d\Omega}{d\omega^{2}}\right|_{\omega=\omega_I} \right)
\mathbf{G}_I = 1 .
\label{eq:G}
%\tag{8}
\end{equation}
Adiabatic approximations yield no frequency-dependence in the matrix of Eq.~(\ref{eq:casida}), but frequency-dependence is at the heart of DTDDFT in capturing double excitations. Here we will consider DTDDFT in a case where the KS single excitations are well-separated, such that a small-matrix approximation (SMA) may be applied that keeps only the diagonal elements of Eq.~(\ref{eq:casida}). 
When one single-excitation $q = i \to a$ %($i$ labeling the occupied and $a$ the unoccupied orbitals) 
mixes with one double-excitation, DTDDFT reduces to dressed SMA (DSMA) where the xc kernel has the diagonal element~\cite{DM23,BM25},~\footnote{The form of the DSMA kernel in Eq.~(\ref{eq:dsma-anya}) combines the ``A'' and ``$0$'' variants of the xc kernel derived in  Ref.~\cite{DM23,DM25}.}:
\begin{align}
f_{\mathrm{XC},\,qq}^{\mathrm{DSMA}_\text{A0}}(\omega)
&= f_{\mathrm{XC},\,qq}^{\mathrm{A}}
+ \frac{\lvert H_{qd}\rvert^{2}}{4\,v_q} \notag \\
&\quad\times \left(
  1 +
  \frac{\bigl(\Omega_q^{\mathrm A} + (H_{dd}-H_{00})\bigr)^{2}}
       {\omega^{2}-\bigl[(H_{dd}-H_{00})^{2}+H_{qd}^{2}\bigr]}
\right).
\label{eq:dsma-anya}
%\tag{9}
\end{align}
Here $f_{\mathrm{XC}}^{\mathrm{A}}$ is a chosen adiabatic xc kernel approximation, upon which DSMA is built, $H_{ij}$  are the matrix elements of the true Hamiltonian involving the KS ground state ($0$), single-excitation ($q$) and double-excitation ($d$), and $\Omega_q^{\mathrm A}$ denotes the adiabatic TDDFT prediction for the single-excitation. 
%In our model,  $\Omega_q^{\mathrm A}$ is well-approximated in SMA. 

 Due to its frequency-dependence, Eq.~(\ref{eq:dsma-anya}) folds in the double-excitation to the single-excitation $q$, shifting the adiabatic prediction for the latter and generating an additional excitation, and Eq.~(\ref{eq:G}) redistributes the KS transition density into that for the resulting states of mixed single and double excitation character. The densities of these states are obtained from the functional derivative of their energies with respect to the external potential, following the procedure in Ref.~\cite{BM25}. Thus, we obtain $E_n, \rho_{n0}, \rho_{nn}$ for these two states. The remaining ingredient for RR-TDDFT is $\rho_{nm}$, the transition density between these states and other excited states. For this, a dressed quadratic response would need to be developed. Instead, here we simply approximate $\rho_{nm}$ along the lines of the auxiliary wavefunction approach of Ref.~\cite{TCR09}, using the eigenvectors $G_I$ to weight the KS determinants. { For the simple model system we study here, the excitations appear in relatively separated multiplet structures, and this approximation is reasonable, as manifest when comparing the resulting couplings to the exact ones. In general, relaxation terms from a proper quadratic response formalism may be more important.  }
Finally, we note that the dressed xc kernel is a correction only to this single excitation, while the other excitations of the system are predicted with adiabatic TDDFT within the usual TDDFT matrix framework.

%This DSMA variant is employed to obtain excitation densities for states with substantial double-excitation character, while the adiabatic exact-exchange approximation is used for states dominated by single-excitation character, following the protocol described in~\cite{BM25}. Within this framework, we also obtain \sout{the generalized intermediate (GI) excitation vectors} {\color{magenta}the eigenvectors $\mathbf G_I$}, which are then used to construct auxiliary wavefunctions and to extract excited-state--to--excited-state transition dipoles. 

%excitation energies and transition dipoles are presented in Tables 1, 2, and 3 for the methods of interest. couplings~\cite{BM25}.

We now demonstrate the capability of RR-TDDFT for dynamics involving double-excitations on a simple model system of two soft-Coulomb interacting electrons in a 1D  anharmonic oscillator:
% \[
% \begin{aligned}
% \label{eq:Harmonic}
% \hat{H}(x_1, x_2) \;=\;& 
% -\tfrac{1}{2}\!\left( \frac{\partial^2}{\partial x_1^2} + \frac{\partial^2}{\partial x_2^2} \right)
% + \frac{1}{\sqrt{(x_1 - x_2)^{2} + \lambda^{2}}}
% \\[0.6ex]
% &\qquad+\frac{1}{2}\bigl(x_1^2 + x_2^2\bigr) + \gamma \bigl(x_1^3 + x_2^3\bigr).
% \end{aligned}
% %\tag{1}
% \]
\ben
\label{eq:Harmonic_short}
\hat{H}(x_1, x_2) = \sum_{i=1,2} \left(-\frac12\partial^2_{x_i}  + \hat v\ext(x_i)\right) + \hat{w}(x_1,x_2)\,,
\een
where $\hat w(x_1,x_2) = \frac{1}{\sqrt{(x_1 - x_2)^{2} + \lambda^{2}}}$ with $\lambda = 2$ being the softening parameter, and $\hat v\ext(x) = \frac12 x^2 + \gamma x^3$ with the parameter $\gamma = 0.05$ that tunes the cubic anharmonicity.
 The anharmonicity is needed to avoid parity preventing access to excited states with an applied laser field of $v^{\rm app}(x,t) = {\mathcal E}(t)x$. 
 %The anharmonic term with the tunable parameter {\color{purple}$\gamma = 0.05$} {\color{purple} \sout{is included to break} breaks} the parity symmetry of the harmonic oscillator eigenstates {\color{purple}allowing the transitions between ... singlet states. { {\it note that we did this because octopus propagations involve a uniform field, i.e. dipole op. If we could propagate with something like $x^2$ we would not need to. Wouldn't write all that but would point to the driving field.} }
With these parameters, the unperturbed system is weakly-interacting enough that the predominant correlation effect is through mixing of a single excitation with a double excitation; this allows us to highlight the ability of RR-TDDFT to describe dynamics accessing these states.   The lowest four exact excitation energies are (in a.u.): $0.9787, 1.8462, 1.9478, 2.7750$ where the
second and third excitations have mixed single and double character, with  $G_2^2$= 0.775  and $G_3^2$= 0.316 indicating their single-excitation components respectively~\cite{DM23}. Adiabatic TDDFT yields only one excitation in the second multiplet, and for our choice of adiabatic exact exchange (AEXX) the three energies in this range are (in a.u.): 0.9788, 1.8848, 2.7900. DSMA$_{\mathrm{A0}}$ built on AEXX (DSMA/AEXX) retrieves the mixing with the KS  double excitation,  yielding four energies in reasonable agreement with the exact (in a.u.): $0.9788, 1.8456, 1.9490, 2.7327$, with the single-excitation components of the second and third states predicted as $G_2^2 =0.6215$ and $G_3^2 = 0.3784$. 
The remaining ingredients for RR-TDDFT are the permanent and transition dipoles; these are given  in Tables~\ref{tab:perm_dipoles_reduced} and~\ref{tab:trans_dipoles_reduced_transposed} respectively. Note that the missing elements for AEXX are those involving the missing state in the second multiplet.

The exact results in the tables were obtained by solving the two-electron Schr\"odinger equation in the \texttt{Octopus}~\cite{octopus,octopus2} code, 
which was also used for the EXX ground state, KS states,  AEXX excitation energies, AEXX ground-to-excited transition dipoles, as well as the TDKS and exact (denoted TDSE) dynamics (see shortly). { A grid-spacing of 1 a.u. and hard-wall box size of 10 a.u. was used.  } The excited-state densities based on Ref.~\cite{BM25}, transition densities based on weighted KS determinants, and the RR-TDDFT dynamics were calculated using an in-house code. 
{ A time-step of 0.03 au was used for both the TDSE and TDKS runs in Octopus using the enforced time-reversal symmetry method, and the RR-TDDFT coefficient dynamics using RK45.}

\begin{table}[htbp]
\caption{Permanent dipole moments, $d_{mm}$ (a.u.) from AEXX and DSMA$_{\mathrm{A0}}$ based on AEXX (DSMA/AEXX),  compared to the exact. }
  \label{tab:perm_dipoles_reduced}
  \centering
  \begin{tabular}{|c|c|c|c|c|c|}
    \hline
     & $d_{00}$ & $d_{11}$ & $d_{22}$ & $d_{33}$ \\
    \hline
    \textbf{ Exact} & -0.1601 & -0.3240 & -0.5038 & -0.4948 \\
    \textbf{AEXX} & -0.1599 & -0.3245 & -0.5084 & -- \\
    \textbf{DSMA/AEXX} & -0.1599 & -0.3159 & -0.5037 & -0.4938 \\
    \hline
  \end{tabular}
\end{table}

%====================== Transition Dipoles ======================
% \begin{table}[htbp]
%   \caption{Transition dipole moments $\mu_{ij}$ (a.u.) from selected methods.}
%   \label{tab:trans_dipoles_reduced}
%   \centering
%   \begin{tabular}{|c|c|c|c|c|}
%     \hline
%     \multicolumn{1}{|c|}{$\mu_{ij}$} & 
%     \multicolumn{1}{c|}{\textbf{Exact}} & 
%     \multicolumn{1}{c|}{\textbf{EXX}} & 
%     \multicolumn{1}{c|}{\textbf{AEXX}} & 
%     \multicolumn{1}{c|}{\textbf{DSMA\_AO}} \\
%     \hline
%     $\mu_{01}$ & -1.007274 & -1.024643 & 1.007296  & 1.007296 \\
%     $\mu_{02}$ & 0.051773  & 0.058203  & -0.060400 & 0.045783 \\
%     $\mu_{03}$ & 0.032548  & --        & --        & 0.034767 \\
%     $\mu_{12}$ & 0.185984  & --        & --        & 0.181309 \\
%     $\mu_{13}$ & -1.418028 & --        & --        & -1.441077 \\
%     $\mu_{23}$ & -0.008978 & --        & --        & 0.005084 \\
%     \hline
%   \end{tabular}
% \end{table}

\begin{table}[htbp]
  \caption{As for Table \ref{tab:perm_dipoles_reduced}, but for the magnitude of the transition dipole moments, $d_{mn}$ (a.u.)}
  \label{tab:trans_dipoles_reduced_transposed}
  \centering
  \begin{tabular}{|c|c|c|c|c|c|c|c|}
    \hline
     & $d_{01}$ & $d_{02}$ & $d_{03}$ & $d_{12}$ & $d_{13}$ & $d_{23}$ \\
    \hline
    \textbf{ Exact} & 1.0072 & 0.0518 & 0.0325 & 0.1860 & 1.4180 & 0.0090 \\
    \textbf{AEXX} & 1.0073 & 0.0604 & -- & 1.029 & -- & -- \\
    \textbf{DSMA/AEXX} & 1.0073 & 0.0458 & 0.0348 & 0.1813 & 1.4411 & 0.0051 \\
    \hline
  \end{tabular}
\end{table}

\begin{figure}
% \RaggedRight
\includegraphics[width=.50\textwidth]{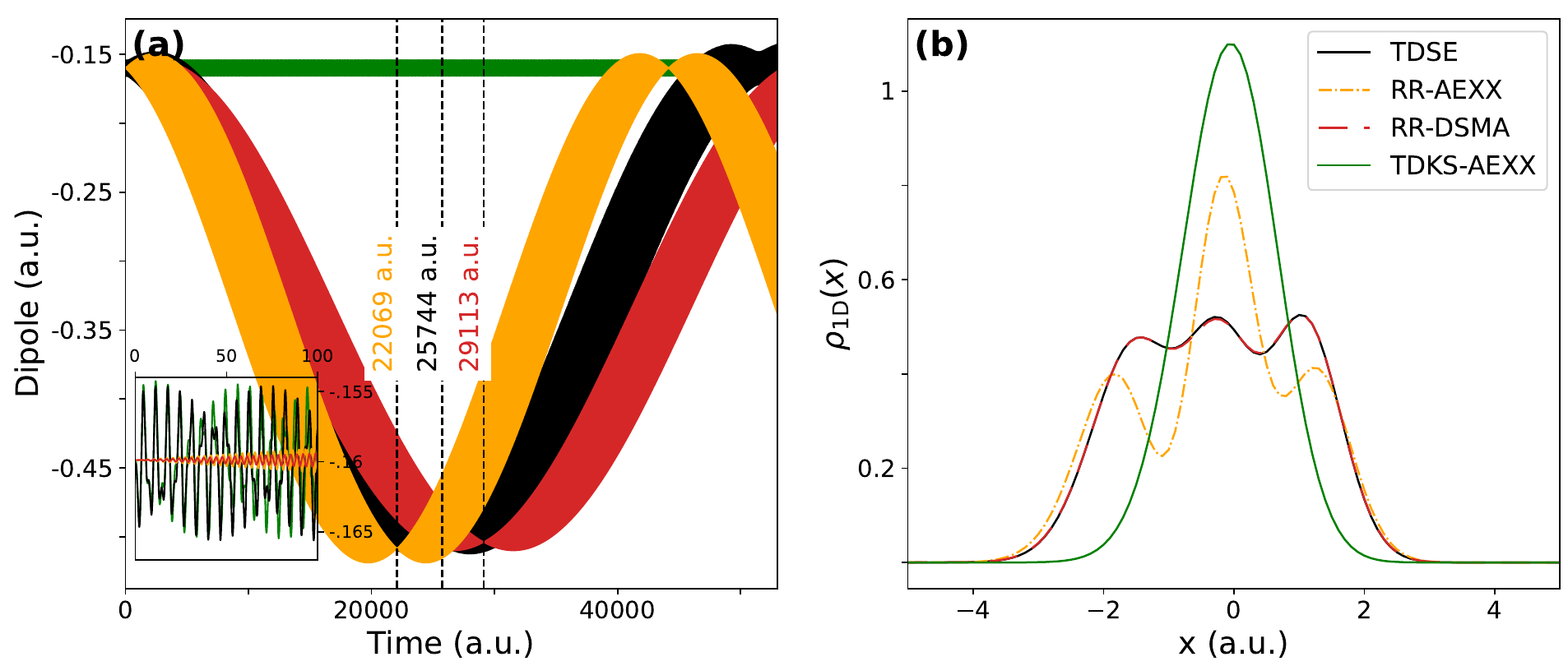}
\caption{Rabi oscillations to the second excited state: a) Dipole predicted by the TDKS propagation with AEXX (green), AEXX used within RR-TDDFT (RR-AEXX, orange), DSMA/AEXX within RR-TDDFT (RR-DSMA, red), and exact (TDSE, black), each driven at their respective resonant frequency. The vertical lines indicate the half-Rabi period for the respective method, and the inset zooms into short times. b) The densities for each of the calculations in panel a, plotted at the respective $T_R/2$. 
 }
\label{fig:Rabi}
\end{figure}

Turning now to dynamics, we drive the system at the frequency of the second excitation, $\omega_2$, to induce a Rabi oscillation, with applied field $
\mathcal{E}(t) = \mathcal{E}_{0}\sin\big(\omega_2\, t\big)\, \hat{\mathbf{x}}
$ where $\mathcal{E}_{0} = 1/(300\sqrt{2})= 0.0023567$~a.u.
After half a Rabi period,
$\frac{T_R}{2} = \frac{\pi}{d_{02} \mathcal{E}_{0}} \approx 25749
$~a.u.,
we expect the population to fully transfer to the excited state. Figure~\ref{fig:Rabi} shows the results of TDKS propagation with AEXX, RR-TDDFT with AEXX, RR-TDDFT with DSMA built on AEXX, all compared to the exact TDSE solution, each driven at their respective resonant $\omega_2$. The TDKS-AEXX dipole strikingly fails to capture the Rabi oscillation, because 
%{ {\it is this part really required?} as expected from the well-known failure of adiabatic TDKS to yield Rabi oscillations, resulting from the reasons discussed earlier: 
ground-state xc is inadequate to describe systems far from a ground state.
On the other hand, when used within RR-TDDFT, AEXX  does yield a Rabi oscillation, and the density does correspond to the density of the AEXX excited state, however because the description of this state lacks the double-excitation component, the density is badly reproduced as evident in Figure~\ref{fig:Rabi}b. RR-TDDFT used with DSMA/AEXX captures the dipole Rabi oscillation  and also accurately reproduces the exact density at $T_R/2$. 
The inset in Figure~\ref{fig:Rabi}a showing very short times, indicates that the error in RR-TDDFT here arises since it was calculated with only two states, however this error is very small on the scale of the overall density change over time. 

%\begin{figure}
%    \centering
%    \includegraphics[width=0.23\textwidth]{figures/comparison_plot.png}
%    \caption{This figure shows the density of each method at their respective half-rabi period {\it merge with Fig 1, see comments there}}
%    \label{fig:omega=.98_1}
%\end{figure}

Next, we give an example where more than two states are involved in the dynamics, and apply a pulse of the form $\mathcal{E}(t) =  \mathcal{E}_0\sin^2\left(\dfrac{\pi t}{2\sigma}\right)\sin(\omega_c t)$ that turns off at $t = 2\sigma$. The driving frequency $\omega_c$, amplitude $\mathcal{E}_0$, and pulse width $\sigma$ are tuned to achieve a coherent evolving superposition of the four lowest states. To isolate errors of the DSMA/AEXX functional approximation from that of truncating our Hilbert space in RR-TDDFT to four states, we further compute the  ingredients for the latter using exact wavefunctions, denoting this truncated TDCI, or TDCI for short.

%We then proceed to reproduce multistate dynamics using RR-TDDFT, marking the first time the method is applied to dynamics involving more than two states. The applied pulse is constructed as a product of an envelope and a carrier component, defined as:
%\begin{equation}
%    \texttt{envelope} =
%    \begin{cases}
%        \sin^2\left(\dfrac{\pi t}{2\sigma}\right), & \text{if } 0 \leq t \leq T_{\mathrm{off}}, \\[6pt]
%        0, & \text{otherwise,}
%    \end{cases}
%\end{equation}
%\begin{equation}
%    \texttt{carrier} = %\sin(\omega_{\pi} t)\, {f_0}.
%\end{equation}
%The total field is then expressed as:
%\begin{equation}
%    \mathcal{E}(t) = \texttt{envelope} %\times \texttt{carrier}.
%\end{equation}
%This pulse shape ensures a smooth turn-on and turn-off within the interval $[0, T_{\mathrm{off}}]$, avoiding discontinuities and emulating a realistic femtosecond laser envelope. 

\begin{figure}
    \centering
    \includegraphics[width=0.5\textwidth]{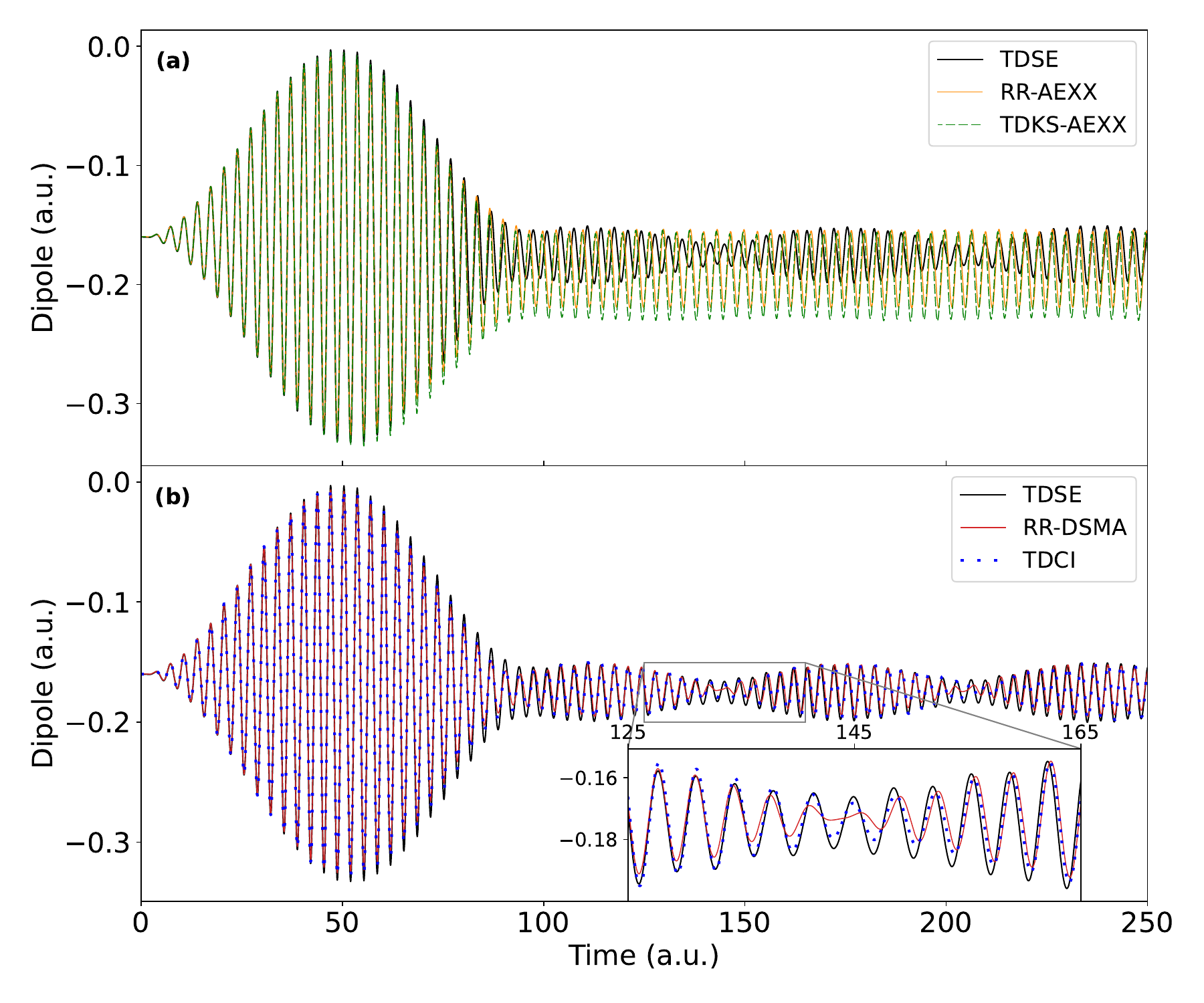}
    \caption{ The dipole moment driven with Pulse 1, predicted by a)  TDKS propagation with AEXX (green, dashed), AEXX used within RR-TDDFT (RR-AEXX, orange), and exact (black), b) DSMA built on AEXX within RR-TDDFT (RR-DSMA, red), truncated TDCI (blue, dotted), and exact (black). }
    \label{fig:omega=1.9}
\end{figure}

Figure~\ref{fig:omega=1.9} shows dynamics resulting from 
a non-resonant driving frequency  in between the second and third excited states of $\omega_c = 1.9$~a.u, strength $\mathcal{E}_0=0.3$~a.u, and width parameter $\sigma = 50$~a.u, denoted Pulse 1.  In the exact evolution, the pulse causes about 4\% of the population (not shown) to leave the ground state, almost equally populating the second and third excited states; during the pulse there is also a small transient population in the lowest excited state.
Both TDKS-AEXX and RR-AEXX give a qualitatively different prediction of the dipole oscillations after about $t = 75$~ a.u., missing the beating and overestimating the average dipole magnitude. 
Unlike the previous case, TDKS-AEXX and RR-AEXX predict  similar dynamics; this is likely because during much of the pulse the external potential dominates over  xc, and because relatively small population leaves the ground state. RR-AEXX yields about twice as much population in its second excited state as the exact, and the purely single-excitation character of the AEXX state results in a wrong density and dipole moment. 
In contrast, RR-DSMA/AEXX accurately captures both the magnitude and the beating pattern. The accuracy of the TDCI confirms that the 4-state truncation is adequate here. The small discrepancy between RR-DSMA/AEXX and exact highlighted in the inset is caused by the small errors in the DSMA predictions of the energies and (transition) densities. 
%A movie of the corresponding densities is provided in the Supplementary Information. 

%Comparing the 
%correctly describing the superposition of states reachedas it includes third excited state and yields corrected excitation energies.
%s shown in Figure 5, the second and third excited states are actively involved in the dynamics and remain partially occupied after the pulse is turned off, giving rise to the characteristic beating pattern. 
%Because AEXX does not have access to the third excited state accurately, it cannot reproduce this behavior.

\begin{figure}
    \centering
    \includegraphics[width=0.5\textwidth]{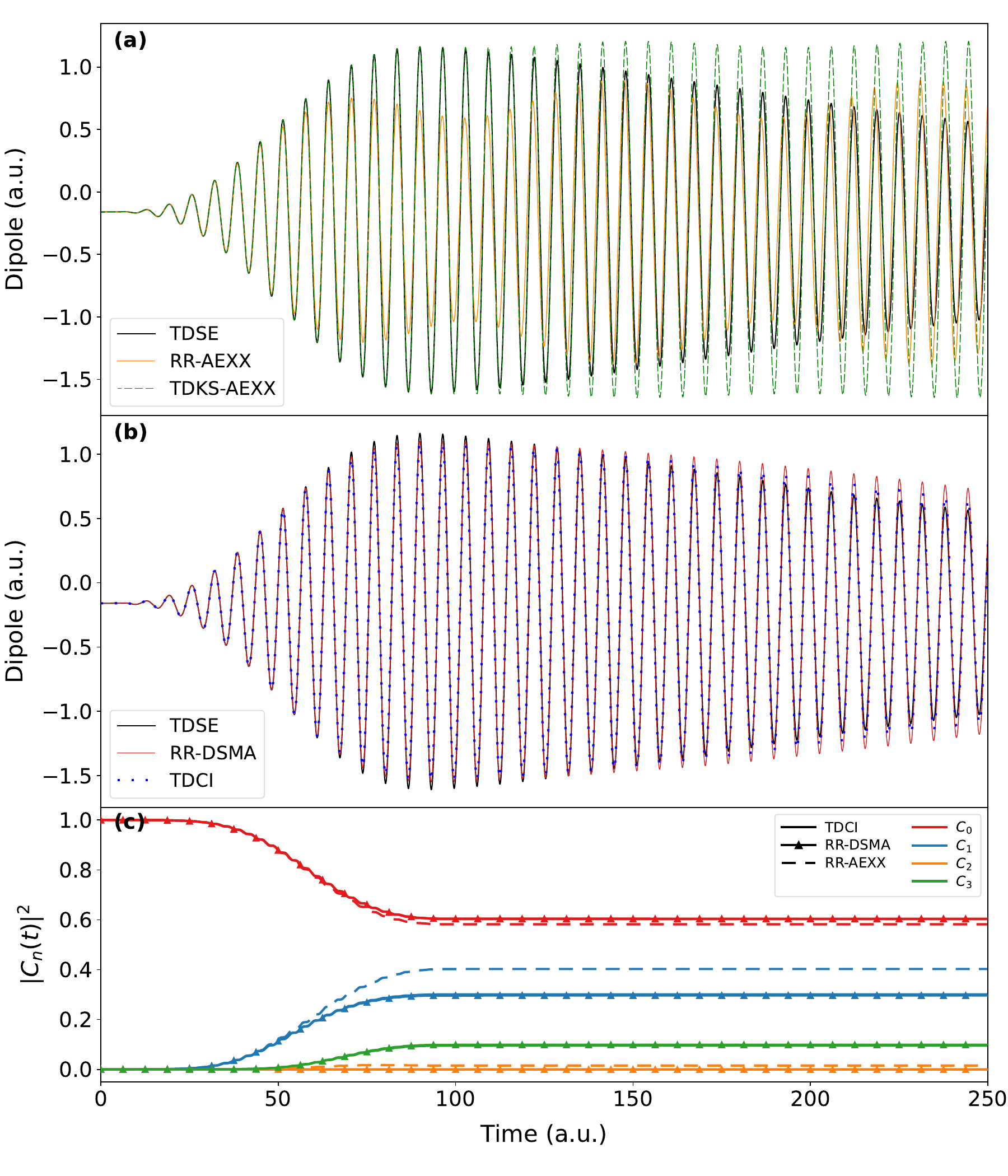}
    \caption{ The dipole moment driven by Pulse 2: a) predicted by TDKS propagation with AEXX (green, dashed), AEXX used within RR-TDDFT (RR-AEXX, orange), and exact (black), b) predicted by DSMA built on AEXX within RR-TDDFT (RR-DSMA, red), truncated TDCI (blue, dotted), and exact (black).
    c) Populations of the first 4 states under Pulse 2.}
    \label{fig:omega=.98}
\end{figure}

Finally, we drive the system at a frequency near the first excited state, with a strength that induces significant population in all three lowest excitations. Choosing $\mathcal{E}_0=0.04$~a.u, $\omega_c= 0.98$~a.u, and $\sigma = 50$~a.u, denoted Pulse 2, Figure~\ref{fig:omega=.98}  shows that after the increase of the dipole during the pulse,  TDKS-AEXX fails to capture the decrease in  magnitude of the exact dipole oscillations over the time period shown.  The decrease manifests the beginning of free-evolution of a superposition of predominantly 4 states (see Fig.~\ref{fig:omega=.98}c), giving modulations to the dominant frequency oscillations $\omega_1$. It is qualitatively captured by RR-AEXX, but the incorrect frequency and dipole moment of the second excited state caused by the lack of double-excitation character gives an incorrect beat frequency. RR-TDDFT with DSMA/AEXX ingredients accurately captures the dynamics.  
The truncated-TDCI and the exact dipoles suggest the small discrepancy in the RR-DSMA/AEXX dipole is both from small populations transferred to higher excited states, and from the small differences in the energies and (transition) density predictions. The RR-DSMA populations shown in Fig.~\ref{fig:omega=.98}c track the exact closely, while RR-AEXX only significantly transfers to the first excited state.  Movies of the corresponding densities under Pulse 1 and Pulse 2 are provided in the Supplementary Information. 

%Movies provided in the Supplementary Information show the density for both the cases in Figs~\ref{fig:omega=1.9} and~\ref{fig:omega=.98}, clearly demonstrating the large qualitative errors in TDKS/AEXX, the smaller but still significant errors in RR-AEXX, and the comparatively accurate predictions of RR-DSMA. \textcolor{orange}{If look at the movie for omega =1.9 I think RR-AEXX is actually worst than TDKS because if we look at the dipole magnitudes the systems dipole moment is changing not drastically and because or system is weakly interacting vexternal domnatates the vh but it is worse in the omega=1.9. However in the on reasonance case of omega=.98 RR-AEXX does a way better job of giving a qualitative decription of the density that TDKS}

%\textcolor{orange}{Movies in the Supplementary Information show the density evolution for the cases in Figs.~\ref{fig:omega=1.9} and~\ref{fig:omega=.98}. They reveal large qualitative errors in TDKS/AEXX, whereas RR-DSMA is comparatively accurate. For the off-resonant case, \(\omega=1.9\), RR-AEXX appears worse than TDKS: the system is weakly interacting and driven off resonance and \(v_{\mathrm{ext}}\) dominates \(v_{\mathrm{H}}\) as evident in the dipole moment changing slightly; in this regime TDKS/AEXX performs better. By contrast, near resonance at \(\omega=0.98\), we leave the perturbative regime and RR-AEXX provides a much more faithful qualitative description of the density than TDKS.}

% \begin{figure}
%     \centering
%     \includegraphics[width=0.5\textwidth]{figures/omega=1.9,sigma=50,f0=.3_dipole(2).pdf}
%     \caption{fig:omega=1.9 {\it see comment for prev fig. Also, maybe we need an inset to show the discrepancy, it's hard to see unless you blow the figure up} }
%     \label{fig:omega=1.9}
% \end{figure}
%\begin{figure}
%    \centering
    %\includegraphics[width=0.5\textwidth{figures/coeffs_only_samecolor.pdf}
%    \caption{fig:omega=1.9 { i think we will eliminate this fgurem or put it in supp matt but maybe dont need}}
%    \label{fig:omega=1.9}
%\end{figure}

% \begin{figure}
%     \centering
%     \includegraphics[width=0.5\textwidth]{figures/omega=.98,f0=.04,sigma=50_dipole(2).pdf}
%     \caption{blah }
%     \label{fig:omega=.98_2}
% \end{figure}

In summary, we have shown how non-perturbative dynamics accessing double excitations can be reliably reproduced in TDDFT by using DTDDFT approximations developed in the response regime within the RR-TDDFT formalism. The idea generalizes:  any functional developments made in the response regime, which has proved to be a more fruitful playground for non-adiabatic functionals than the real-time domain, can be exploited for non-equilibrium dynamics via RR-TDDFT. The ingredients are excitation energies and transition densities out of the ground state, obtained from poles and residues of the linear density response function (equally, eigenvalues and eigenvectors of the matrix equation), excited-state densities from the potential-functional-derivative of the energies, and transition-densities between excited states from quadratic response, or approximated by auxiliary wavefunctions as done in the present work. 
As discussed in Ref.~\cite{DBM24}, RR-TDDFT requires calculating $M(M-1)$ couplings, where $M$ is the number of states anticipated to be involved in the dynamics.  But the response quantities need be computed once and for all, and then can be used for any applied fields~\cite{LRG25}. { We emphasize that RR-TDDFT will only be as good as the TDDFT description of the response spectra. For example, if the dynamics accesses continuum states, then an appropriate and large enough basis would be needed to compute the energies and response quantities; the accuracy of the dynamics from RR-TDDFT will directly relate to their accuracy, and further, unphysical reflections from reaching any boundary need to be sensibly treated. We note that analogous basis-set  and boundary concerns appear in the traditional TDKS approach.} 

{Concerning the observables: as in the traditional TDKS approach, observables directly related to the density can be extracted without further approximation, but other observables such as kinetic energy spectra, double ionization yields, momentum densities, need in principle to be expressed through an additional functional of the density. In the traditional TDKS approach, often the TDKS time-dependent determinant is treated as an approximation of the true wavefunction for this particular purpose, a somewhat uncontrolled approximation.  A similar approximation could be made in RR-TDDFT, treating the configuration-state functions constructed from the ground-state KS determinant and its excitations as approximations to the static many-body wavefunctions.}

Future work will apply RR-TDDFT with DTDDFT to dynamics involving double excitations in a range of real molecules (e.g., Refs.~\cite{IL08,MIFS14}), establish quadratic response within the DTDDFT approach for excited-excited state couplings, and explore other non-adiabatic functionals developed in the response regime (e.g.~\cite{VK96,VUC97} ) that can be now used for non-equilibrium dynamics via RR-TDDFT.

\acknowledgments{Financial support from the National Science Foundation Award CHE-2154829 (DR and NTM) and the Department of
Energy, Office of Basic Energy Sciences, Division of
Chemical Sciences, Geosciences and Biosciences under
Award No. DE‐SC0024496 (AB) are gratefully acknowledged. }

\bibliography{ref.bib}

@article{SVHA02,
    author = {Salek, Pawel and Vahtras, Olav and Helgaker, Trygve and Ågren, Hans},
    title = {Density-functional theory of linear and nonlinear time-dependent molecular properties},
    journal = {The Journal of Chemical Physics},
    volume = {117},
    number = {21},
    pages = {9630-9645},
    year = {2002},
    month = {12},
    issn = {0021-9606},
    doi = {10.1063/1.1516805},
    url = {https://doi.org/10.1063/1.1516805},
    eprint = {https://pubs.aip.org/aip/jcp/article-pdf/117/21/9630/19225236/9630\_1\_online.pdf},
}

@Inbook{PF18,
author="Parker, Shane M.
and Furche, Filipp",
editor="W{\'o}jcik, Marek J.
and Nakatsuji, Hiroshi
and Kirtman, Bernard
and Ozaki, Yukihiro",
title="Response Theory and Molecular Properties",
bookTitle="Frontiers of Quantum Chemistry",
year="2018",
publisher="Springer Singapore",
address="Singapore",
pages="69--86",
isbn="978-981-10-5651-2",
doi="10.1007/978-981-10-5651-2_4",
url="https://doi.org/10.1007/978-981-10-5651-2_4"
}

@article{PSLP21,
author = {Park, Woojin and Shen, Jun and Lee, Seunghoon and Piecuch, Piotr and Filatov, Michael and Choi, Cheol Ho},
title = {Internal Conversion between Bright (11Bu+) and Dark (21Ag–) States in s-trans-Butadiene and s-trans-Hexatriene},
journal = {The Journal of Physical Chemistry Letters},
volume = {12},
number = {39},
pages = {9720-9729},
year = {2021},
doi = {10.1021/acs.jpclett.1c02707},
    note ={PMID: 34590847},

URL = { 
    
        https://doi.org/10.1021/acs.jpclett.1c02707
    
    

},
eprint = { 
    
        https://doi.org/10.1021/acs.jpclett.1c02707
    
    

}

}

@article{DBM24,
  title = {Reformulation of Time-Dependent Density Functional Theory for Nonperturbative Dynamics: The Rabi Oscillation Problem Resolved},
  author = {Dar, Davood B. and Baranova, Anna and Maitra, Neepa T.},
  journal = {Phys. Rev. Lett.},
  volume = {133},
  issue = {9},
  pages = {096401},
  numpages = {7},
  year = {2024},
  month = {Aug},
  publisher = {American Physical Society},
  doi = {10.1103/PhysRevLett.133.096401},
  url = {https://link.aps.org/doi/10.1103/PhysRevLett.133.096401}
}

@Article{octopus,
author ="Andrade, Xavier and Strubbe, David and De Giovannini, Umberto and Larsen, Ask Hjorth and Oliveira, Micael J. T. and Alberdi-Rodriguez, Joseba and Varas, Alejandro and Theophilou, Iris and Helbig, Nicole and Verstraete, Matthieu J. and Stella, Lorenzo and Nogueira, Fernando and Aspuru-Guzik, Alán and Castro, Alberto and Marques, Miguel A. L. and Rubio, Angel",
title  ="Real-space grids and the Octopus code as tools for the development of new simulation approaches for electronic systems",
journal  ="Phys. Chem. Chem. Phys.",
year  ="2015",
volume  ="17",
issue  ="47",
pages  ="31371-31396",
publisher  ="The Royal Society of Chemistry",
doi  ="10.1039/C5CP00351B",
url  ="http://dx.doi.org/10.1039/C5CP00351B"}

@article{F01,
author = {Furche,Filipp },
title = {On the density matrix based approach to time-dependent density functional response theory},
journal = {The Journal of Chemical Physics},
volume = {114},
number = {14},
pages = {5982-5992},
year = {2001}
}

@article{FA02,
author = {Furche,Filipp  and Ahlrichs,Reinhart },
title = {Adiabatic time-dependent density functional methods for excited state properties},
journal = {The Journal of Chemical Physics},
volume = {117},
number = {16},
pages = {7433-7447},
year = {2002}
}

@article{DLM22,
author = {Dar,Davood  and Lacombe,Lionel  and Maitra,Neepa T. },
title = {The exact exchange–correlation potential in time-dependent density functional theory: Choreographing electrons with steps and peaks},
journal = {Chem. Phys. Rev.},
volume = {3},
number = {3},
pages = {031307},
year = {2022},
doi = {10.1063/5.0096627},}

@article{LM23,
	author = {Lacombe, Lionel and Maitra, Neepa T.},
	journal = {npj Comput. Mater.},
	number = {1},
	pages = {124},
	title = {Non-adiabatic approximations in time-dependent density functional theory: progress and prospects},
	volume = {9},
	year = {2023}}

@article{M16,
author = {Maitra,Neepa T. },
title = {Perspective: Fundamental aspects of time-dependent density functional theory},
journal = {The Journal of Chemical Physics},
volume = {144},
number = {22},
pages = {220901},
year = {2016},
doi = {10.1063/1.4953039},

URL = { 
        https://doi.org/10.1063/1.4953039
    
},
eprint = { 
        https://doi.org/10.1063/1.4953039
    
}

}

@article{M17,
	doi = {10.1088/1361-648x/aa836e},
	url = {https://doi.org/10.1088/1361-648x/aa836e},
	year = 2017,
	month = {sep},
	publisher = {{IOP} Publishing},
	volume = {29},
	number = {42},
	pages = {423001},
	author = {Neepa T Maitra},
	title = {Charge transfer in time-dependent density functional theory},
	journal = {Journal of Physics: Condensed Matter}
}

@article {MW09,
author = {Mazur, Grzegorz and W{\l}odarczyk, Radoslaw},
title = {Application of the dressed time-dependent density functional theory for the excited states of linear polyenes},
journal = {J. Comput. Chem.},
volume = {30},
number = {5},
publisher = {Wiley Subscription Services, Inc., A Wiley Company},
issn = {1096-987X},
pages = {811--817},
year = {2009},
}

@article{C05,
   author = "Casida, Mark E.",
   title = "Propagator corrections to adiabatic time-dependent density-functional theory linear response theory",
   journal = "J. Chem. Phys.",
   year = "2005",
   volume = "122",
   number = "5", 
   eid = 054111,
   pages = "",

}

@article{RSBS09,
   author = "Romaniello, P. and Sangalli, D. and Berger, J. A. and Sottile, F. and Molinari, L. G. and Reining, L. and Onida, G.",
   title = "Double excitations in finite systems",
   journal = "J. Chem. Phys.",
   year = "2009",
   volume = "130",
   number = "4", 
   eid = 044108,
   pages = "",

}

@article{MZCB04,
   author = "Maitra, Neepa T. and Zhang, Fan and Cave, Robert J. and Burke, Kieron",
   title = "Double excitations within time-dependent density functional theory linear response",
   journal = "J. Chem. Phys.",
   year = "2004",
   volume = "120",
   number = "13", 
   pages = "5932-5937",
}

@Article{TH00,
author ="Tozer, David J. and Handy, Nicholas C.",
title  ="On the determination of excitation energies using density functional theory",
journal  ="Phys. Chem. Chem. Phys.",
year  ="2000",
volume  ="2",
issue  ="10",
pages  ="2117-2121",
publisher  ="The Royal Society of Chemistry",

}

@article{GPG00,
  title={Molecular excitation energies from time-dependent density functional theory},
  author={Grabo, T and Petersilka, M and Gross, EKU},
  journal={Journal of Molecular Structure: THEOCHEM},
  volume={501},
  pages={353--367},
  year={2000},
  publisher={Elsevier}
}

@Incollection{C96,
author="Casida, Mark E.",
editor="J. M. Seminario",
title="Time-Dependent Density Functional Response Theory of Molecular Systems: Theory, Computational Methods, and Functionals",
booktitle="Recent Developments and Applications of Modern Density Functional Theory", 
year="1996",
publisher="Elsevier",
address="Amsterdam",
pages="391",
}

@Incollection{C95,
title="Time-dependent density functional response theory for molecules",
editor="D.P. Chong",
  booktitle="Recent Advances in Density Functional Methods, Part I",
  author="Casida, MK",
  year="1995",
  publisher="World Scientific, Singapore"
}

@article{FLSM15,
  title = {Time-Resolved Spectroscopy in Time-Dependent Density Functional Theory: An Exact Condition},
  author = {Fuks, Johanna I. and Luo, Kai and Sandoval, Ernesto D. and Maitra, Neepa T.},
  journal = {Phys. Rev. Lett.},
  volume = {114},
  issue = {18},
  pages = {183002},
  numpages = {6},
  year = {2015},
  month = {May},
  publisher = {American Physical Society},

}

@article{VUC97,
  title = {Time-Dependent Density Functional Theory Beyond the Adiabatic Local Density Approximation},
  author = {Vignale, G. and Ullrich, C. A. and Conti, S.},
  journal = {Phys. Rev. Lett.},
  volume = {79},
  issue = {24},
  pages = {4878--4881},
  numpages = {0},
  year = {1997},
  month = {Dec},
  publisher = {American Physical Society},

}

@article{VK96,
  title = {Current-Dependent Exchange-Correlation Potential for Dynamical Linear Response Theory},
  author = {Vignale, G. and Kohn, Walter},
  journal = {Phys. Rev. Lett.},
  volume = {77},
  issue = {10},
  pages = {2037--2040},
  numpages = {0},
  year = {1996},
  month = {Sep},
  publisher = {American Physical Society},

}

@article{TAHR99,
author = { David J.   Tozer  and  Roger D.   Amos  and  Nicholas C.   Handy  and  Bjorn O.   Roos  and  Luis   Serrano-Andres},
title = {Does density functional theory contribute to the understanding of excited states of unsaturated organic compounds?},
journal = {Mol. Phys.},
volume = {97},
number = {7},
pages = {859-868},
year = {1999},
}

@article{TCR09,
   author = "Tavernelli, Ivano and Curchod, Basile F. E. and Rothlisberger, Ursula",
   title = "On nonadiabatic coupling vectors in time-dependent density functional theory",
   journal = "J. Chem. Phys.",
   year = "2009",
   volume = "131",
   number = "19", 
   eid = 196101,
   pages = "",
}

@article{LFSEM14,
	Author = {Luo, Kai and Fuks, Johanna I and Sandoval, Ernesto D and Elliott, Peter and Maitra, Neepa T},
	Journal = {J. Chem. Phys},
	Number = {18},
	Pages = {18A515},
	Publisher = {AIP Publishing},
	Title = {Kinetic and interaction components of the exact time-dependent correlation potential},
	Volume = {140},
	Year = {2014}}

@article{RG84,
	Author = {Runge, Erich and Gross, E. K. U.},
	Doi = {10.1103/PhysRevLett.52.997},
	Issue = {12},
	Journal = {Phys. Rev. Lett.},
	Month = {Mar},
	Numpages = {0},
	Pages = {997--1000},
	Publisher = {American Physical Society},
	Title = {Density-Functional Theory for Time-Dependent Systems},
	Url = {http://link.aps.org/doi/10.1103/PhysRevLett.52.997},
	Volume = {52},
	Year = {1984}}

@book{Carstenbook,
	Author = {Ullrich, Carsten A},
	Publisher = {Oxford University Press},
	Title = {Time-dependent density-functional theory: concepts and applications},
	Year = {2011}}

@article{PGG96,
	Author = {Petersilka, M. and Gossmann, U. J. and Gross, E. K. U.},
	Doi = {10.1103/PhysRevLett.76.1212},
	Issue = {8},
	Journal = {Phys. Rev. Lett.},
	Month = {Feb},
	Numpages = {0},
	Pages = {1212--1215},
	Publisher = {American Physical Society},
	Title = {Excitation Energies from Time-Dependent Density-Functional Theory},
	Url = {http://link.aps.org/doi/10.1103/PhysRevLett.76.1212},
	Volume = {76},
	Year = {1996}}

@article{EFRM12,
	Author = {Elliott, P. and Fuks, J. I. and Rubio, A. and Maitra, N. T.},
	Doi = {10.1103/PhysRevLett.109.266404},
	Issue = {26},
	Journal = {Phys. Rev. Lett.},
	Month = {Dec},
	Numpages = {5},
	Pages = {266404},
	Publisher = {American Physical Society},
	Title = {Universal Dynamical Steps in the Exact Time-Dependent Exchange-Correlation Potential},
	Url = {http://link.aps.org/doi/10.1103/PhysRevLett.109.266404},
	Volume = {109},
	Year = {2012}}

@article{FERM13,
	Author = {Fuks, Johanna I and Elliott, P and Rubio, A and Maitra, Neepa T},
	Journal = {J. Phys. Chem. Lett.},
	Number = {5},
	Pages = {735--739},
	Publisher = {ACS Publications},
	Title = {Dynamics of Charge-Transfer Processes with Time-Dependent Density Functional Theory},
	Volume = {4},
	Year = {2013}}

@article{FHTR11,
	Author = {Fuks, Johanna I and Helbig, N and Tokatly, IV and Rubio, A},
	Journal = {Phys. Rev. B},
	Number = {7},
	Pages = {075107},
	Publisher = {APS},
	Title = {Nonlinear phenomena in time-dependent density-functional theory: What Rabi oscillations can teach us},
	Volume = {84},
	Year = {2011}}

@article{RN11,
	Author = {Raghunathan, Shampa and Nest, Mathias},
	Doi = {10.1021/ct200270t},
	Journal = {J. Chem.  Theory and Comput.},
	Number = {8},
	Pages = {2492-2497},
	Title = {Critical Examination of Explicitly Time-Dependent Density Functional Theory for Coherent Control of Dipole Switching},
	Url = {http://dx.doi.org/10.1021/ct200270t},
	Volume = {7},
	Year = {2011}}

@article{RN12c,
	Author = {Raghunathan, Shampa and Nest, Mathias},
	Journal = {J. Chem. Theory and Comput.},
	Number = {3},
	Pages = {806--809},
	Publisher = {ACS Publications},
	Title = {The Lack of Resonance Problem in Coherent Control with Real-Time Time-Dependent Density Functional Theory},
	Volume = {8},
	Year = {2012}}

@article{RB09,
	Author = {Ruggenthaler, M. and Bauer, D.},
	Doi = {10.1103/PhysRevLett.102.233001},
	Issue = {23},
	Journal = {Phys. Rev. Lett.},
	Month = {Jun},
	Numpages = {4},
	Pages = {233001},
	Publisher = {American Physical Society},
	Title = {Rabi Oscillations and Few-Level Approximations in Time-Dependent Density Functional Theory},
	Url = {http://link.aps.org/doi/10.1103/PhysRevLett.102.233001},
	Volume = {102},
	Year = {2009}}

@article{HTPI14,
	Author = {Bradley F. Habenicht and Noriyuki P. Tani and Makenzie R. Provorse and Christine M. Isborn},
	Journal = {J. Chem. Phys.},
	Pages = {184112},
	Volume = {141},
	Year = {2014},
	}

@article {octopus2,
author = {Castro, Alberto and Appel, Heiko and Oliveira, Micael and Rozzi, Carlo A. and Andrade, Xavier and Lorenzen, Florian and Marques, M. A. L. and Gross, E. K. U. and Rubio, Angel},
title = {octopus: a tool for the application of time-dependent density functional theory},
journal = {physica status solidi (b)},
volume = {243},
number = {11},
publisher = {WILEY-VCH Verlag},
issn = {1521-3951},
url = {http://dx.doi.org/10.1002/pssb.200642067},
doi = {10.1002/pssb.200642067},
pages = {2465--2488},
year = {2006},
}

@article{DM23,
  title={Oscillator strengths and excited-state couplings for double excitations in time-dependent density functional theory},
  author={Dar, Davood B and Maitra, Neepa T},
  journal={The Journal of Chemical Physics},
  volume={159},
  number={21},
  year={2023},
  publisher={AIP Publishing}
}

@article{LM21,
  title={Minimizing the time-dependent density functional error in ehrenfest dynamics},
  author={Lacombe, Lionel and Maitra, Neepa T},
  journal={The Journal of Physical Chemistry Letters},
  volume={12},
  number={35},
  pages={8554--8559},
  year={2021},
  publisher={ACS Publications}
}

@article{DM25,
  author       = {Dar, Davood B. and Maitra, Neepa T.},
  title        = {Capturing the Elusive Curve‐Crossing in Low‐Lying States of Butadiene with Dressed TDDFT},
  journal      = {The Journal of Physical Chemistry Letters},
  volume       = {16},
  number       = {3},
  pages        = {703--709},
  year         = {2025},
  doi          = {10.1021/acs.jpclett.4c03167},
}

@article{BM25,
  title = {Excited-{{State Densities}} from {{Time-Dependent Density Functional Response Theory}}},
  author = {Baranova, Anna and Maitra, Neepa T.},
  year = {2025},
  month = oct,
  journal = {Journal of Chemical Theory and Computation},
  publisher = {American Chemical Society},
  issn = {1549-9618},
  doi = {10.1021/acs.jctc.5c00909},
  urldate = {2025-10-13},
}

@article{IL08,
  title = {Modeling the Doubly Excited State with Time-Dependent {{Hartree}}--{{Fock}} and Density Functional Theories},
  author = {Isborn, Christine M. and Li, Xiaosong},
  year = {2008},
  month = nov,
  journal = {J. Chem. Phys.},
  volume = {129},
  number = {20},
  pages = {204107},
  issn = {0021-9606},
  doi = {10.1063/1.3020336},
  urldate = {2025-10-13},
}

@misc{LRG25,
  title = {Electron Charge Dynamics and Charge Separation: {{A}} Response Theory Approach},
  shorttitle = {Electron Charge Dynamics and Charge Separation},
  author = {Lacombe, Lionel and Reining, Lucia and Gorelov, Vitaly},
  year = {2025},
  month = aug,
  number = {arXiv:2508.14551},
  eprint = {2508.14551},
  primaryclass = {cond-mat},
  publisher = {arXiv},
  doi = {10.48550/arXiv.2508.14551},
  urldate = {2025-10-14},
  archiveprefix = {arXiv}
}

@article{MIFS14,
  title = {Ultrafast {{Relaxation Dynamics}} in Trans-1,3-{{Butadiene Studied}} by {{Time-Resolved Photoelectron Spectroscopy}} with {{High Harmonic Pulses}}},
  author = {Makida, Ayumu and Igarashi, Hironori and Fujiwara, Takehisa and Sekikawa, Taro and Harabuchi, Yu and Taketsugu, Tetsuya},
  year = {2014},
  month = may,
  journal = {J. Phys. Chem. Lett.},
  volume = {5},
  number = {10},
  pages = {1760--1765},
  publisher = {American Chemical Society},
  doi = {10.1021/jz5003567},
  urldate = {2025-10-14}
}

@article{SS11b,
  title = {{{TD-CI Simulation}} of the {{Electronic Optical Response}} of {{Molecules}} in {{Intense Fields II}}: {{Comparison}} of {{DFT Functionals}} and {{EOM-CCSD}}},
  shorttitle = {{{TD-CI Simulation}} of the {{Electronic Optical Response}} of {{Molecules}} in {{Intense Fields II}}},
  author = {Sonk, Jason A. and Schlegel, H. Bernhard},
  year = 2011,
  month = oct,
  journal = {The Journal of Physical Chemistry A},
  volume = {115},
  number = {42},
  pages = {11832--11840},
  publisher = {American Chemical Society},
  issn = {1089-5639},
  doi = {10.1021/jp206437s},
  urldate = {2026-01-11}
}

@article{CL21,
  title = {Time-Dependent Ab Initio Approaches for High-Harmonic Generation Spectroscopy},
  author = {Coccia, Emanuele and Luppi, Eleonora},
  year = 2021,
  month = nov,
  journal = {Journal of Physics: Condensed Matter},
  volume = {34},
  number = {7},
  pages = {073001},
  publisher = {IOP Publishing},
  issn = {0953-8984},
  doi = {10.1088/1361-648X/ac3608},
  urldate = {2026-01-11},
  langid = {english}
}

@article{MLR14b,
  title = {Charge Migration in the Bifunctional {{PENNA}} Cation Induced and Probed by Ultrafast Ionization: A Dynamical Study},
  shorttitle = {Charge Migration in the Bifunctional {{PENNA}} Cation Induced and Probed by Ultrafast Ionization},
  author = {Mignolet, B and Levine, R D and Remacle, F},
  year = 2014,
  month = jun,
  journal = {Journal of Physics B: Atomic, Molecular and Optical Physics},
  volume = {47},
  number = {12},
  pages = {124011},
  publisher = {IOP Publishing},
  issn = {0953-4075},
  doi = {10.1088/0953-4075/47/12/124011},
  urldate = {2026-01-11},
  langid = {english}
}
\end{document}

\title{Supplementary Information for: Harnessing dressed time-dependent density functional theory for the non-perturbative regime: Electron dynamics with double excitations
}
\author{Dhyey Ray}
\affiliation{Department of Physics, Rutgers University, Newark 07102, New Jersey USA}
%\affiliation{Department of Biomedical Engineering, Rutgers University, Piscataway, New Jersey USA}
\author{Anna Baranova}
\affiliation{Department of Physics, Rutgers University, Newark 07102, New Jersey USA}
\author{Davood B. Dar}
\affiliation{Department of Physics, Rutgers University, Newark 07102, New Jersey USA}
\affiliation{Department of Physical and Environmental Sciences University of Toronto, Canada}
\author{Neepa T. Maitra}
\affiliation{Department of Physics, Rutgers University, Newark 07102, New Jersey USA}
\email{neepa.maitra@rutgers.edu}

\date{\today}
\maketitle
\vspace{-7em}

\section{Movies}
\begin{figure}[H]
    \centering
    \includegraphics[width=.50\textwidth]{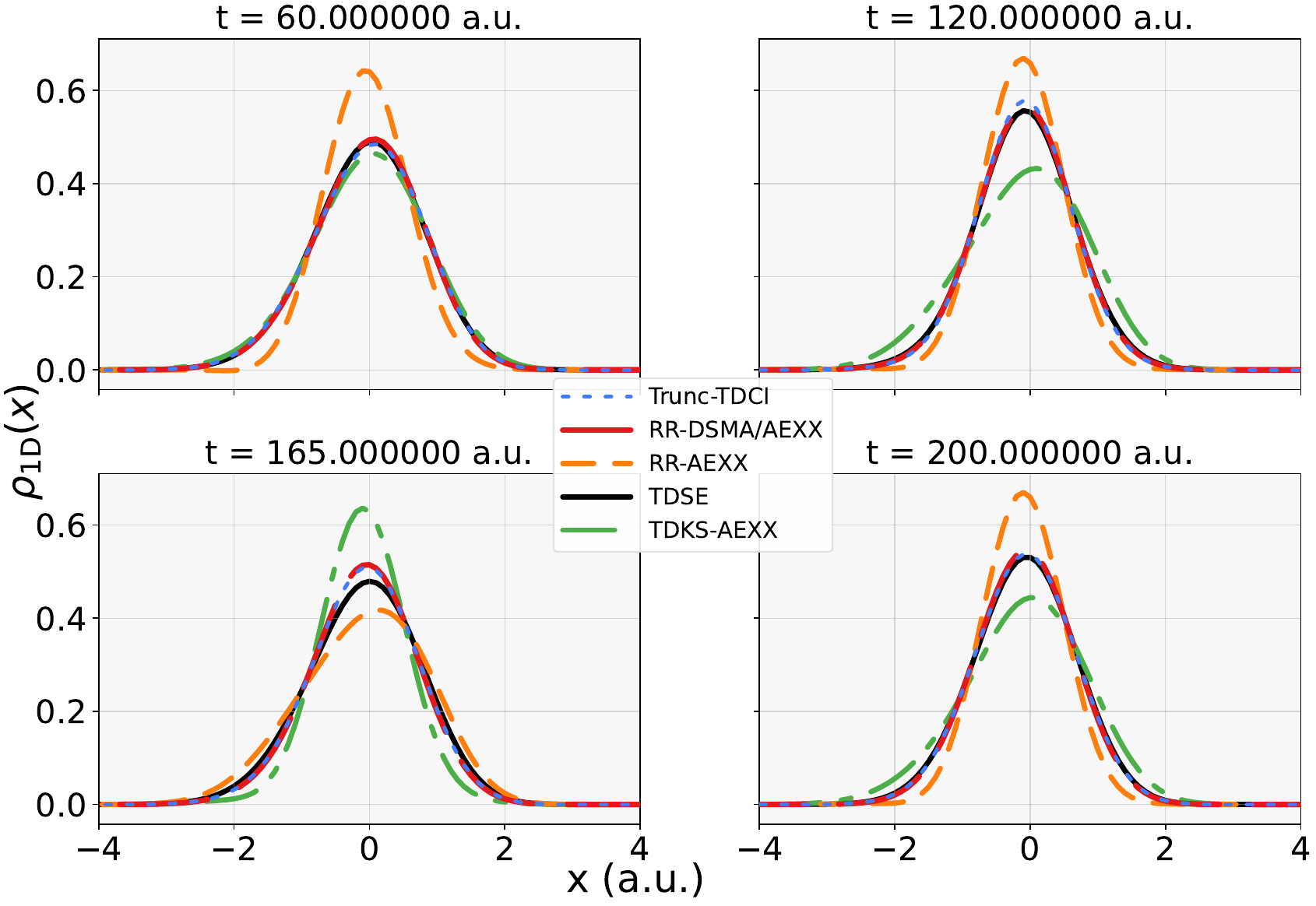}
    \caption{Snapshots from Movie~(i) showing the density driven by Pulse~1 at four different times, predicted by truncated TDCI (blue, dashed), RR-DSMA/AEXX (red), RR-AEXX (orange, dashed), TDKS-AEXX (green, dashed), and exact TDSE (black).}
    \label{fig:Rabi1}
\end{figure}

\begin{figure}[H]
    \centering
    \includegraphics[width=.50\textwidth]{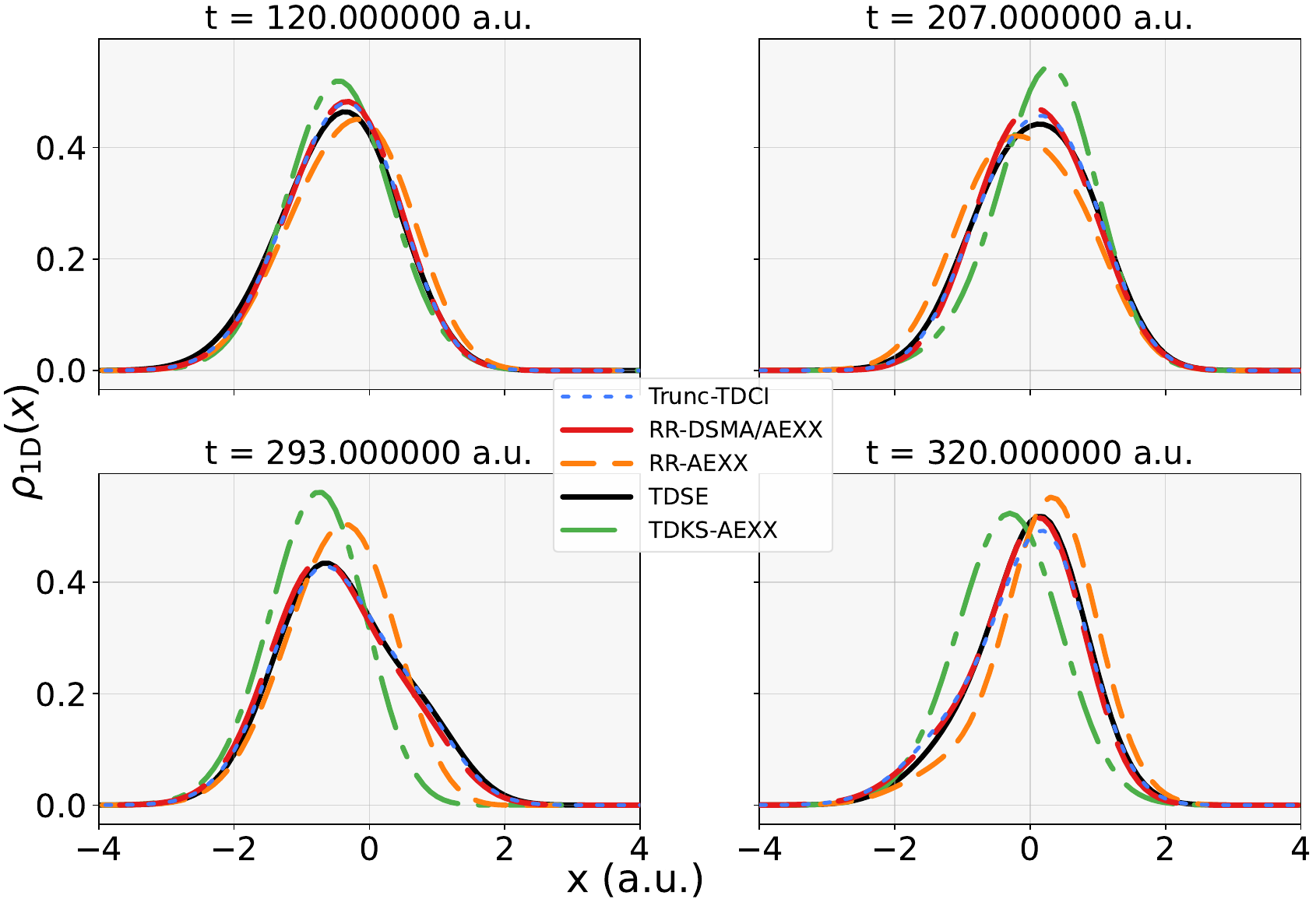}
    \caption{Snapshots from Movie~(ii) showing the density driven by Pulse~2 at four different times, predicted by truncated TDCI (blue, dashed), RR-DSMA/AEXX (red), RR-AEXX (orange, dashed), TDKS-AEXX (green, dashed), and exact TDSE (black).}
    \label{fig:Rabi2}
\end{figure}

\vspace{-0.4\baselineskip}
We provide two movies illustrating the density evolution of the system defined by the Hamiltonian in Eq.~7 under Pulse 1 and Pulse 2 defined in the main text. Snapshots from each movie are provided here in Figs.~\ref{fig:Rabi1}–\ref{fig:Rabi2} respectively. 

\begin{itemize}
    \item[(i)] \textbf{Movie f00p3\_O1p9\_S50.mp4} shows the dynamics under Pulse~1 ($\omega_c = 1.9$~a.u., $\mathcal{E}_0 = 0.3$~a.u., $\sigma = 50$~a.u.). The plot compares the density evolution predicted by the exact TDSE (black), truncated TDCI (blue, dashed), RR-DSMA/AEXX (red), RR-AEXX (orange, dashed), and TDKS-AEXX (green, dashed). The density undergoes small oscillations, with largely a breathing motion, remaining centered near the initial ground state profile, with the performance of the different methods similar to that reflected in the dipole of the main text.
    Figure~\ref{fig:Rabi1} shows four representative snapshots from this movie.

    \item[(ii)] \textbf{Movie f00p04\_Op98\_S50.mp4} shows the dynamics under Pulse~2 ($\omega_c = 0.98$~a.u., $\mathcal{E}_0 = 0.04$~a.u., $\sigma = 50$~a.u.). The same color conventions are used as in Movie~(i). The larger population transfer out of the ground state leads to larger density oscillations (breathing and sloshing). 
    Figure~\ref{fig:Rabi2} shows four representative snapshots from this movie.
\end{itemize}

